\documentclass[a4paper,prd,
twocolumn,nofootinbib,tightenlines,showpacs,preprintnumbers,amsmath,amssymb]{revtex4-1}

\usepackage{times}
\usepackage{amssymb,amsbsy,amsmath,amsfonts}
\usepackage{graphicx}
\usepackage{mathrsfs}
\usepackage{esint}
\usepackage{nicefrac}
\usepackage[colorlinks,citecolor=blue,linktoc=all,linkcolor=cyan]{hyperref} 
\usepackage{upgreek}
\usepackage{multirow}
\usepackage{rotating}
\usepackage{capt-of}
\usepackage{diagbox}

\def\beq{\begin{equation}}
\def\eeq{\end{equation}}
\def\bea{\begin{eqnarray}}
\def\eea{\end{eqnarray}}
\def\beqa{\begin{equation}\begin{array}{l}}
\def\eeqa{\end{array}\end{equation}}

\def\eqlab#1{\label{eq:#1}}
\def\figlab#1{\label{fig:#1}}

\def\eref#1{(\ref{eq:#1})}
\def\eqref#1{eq.~(\ref{eq:#1})}
\def\Eqref#1{Eq.~(\ref{eq:#1})}

\def\Figref#1{Fig.~\ref{fig:#1}}

\def\Secref#1{Sec.~\ref{sec:#1}}

\def\quarter{\mbox{\small{$\frac{1}{4}$}}}

\def\barr{\left(\begin{array}{c}}
\def\earr{\end{array}\right)}
\def\bmat{\left(\begin{array}{cc}}
\def\emat{\end{array}\right)}
\def\al{\alpha}

\def\ga{\gamma} \def\Ga{{\it\Gamma}}
\def\de{\delta} \def\De{\Delta}
\def\veps{\varepsilon}  

\def\la{\lambda} \def\La{{\it\Lambda}}

\def\si{\sigma} \def\Si{{\it\Sigma}}
\def\th{\theta}

\def\dd{{\rm d}}

\def\nn{\nonumber}

\def\ol#1{\overline{#1}}

\DeclareMathOperator\im{Im}
\DeclareMathOperator\re{Re}

\def\beps{\boldsymbol{\varepsilon}}

\def\bsigma{\boldsymbol{\sigma}}

\begin{document}
\preprint{MITP/16-001}
\title {Evaluation of the forward Compton scattering off protons: II. Spin-dependent amplitude and observables}
\author{Oleksii Gryniuk}
\author{Franziska Hagelstein}
\author{Vladimir Pascalutsa}
\affiliation{Institut f\"ur Kernphysik and  PRISMA Cluster of Excellence, Johannes
Gutenberg-Universit\"at Mainz, D-55128 Mainz, Germany}
\begin{abstract}
The forward Compton scattering off the proton is determined
by substituting the empirical total photoabsorption cross sections into
dispersive sum rules. 
In addition to the spin-independent amplitude evaluated previously [Phys.\ Rev.\ D {\bf 92}, 074031 (2015)], 
we obtain the spin-dependent
amplitude over a broad energy range.
The two amplitudes contain all the information about this process, and we, hence, can reconstruct the nonvanishing observables of the proton Compton scattering in the forward kinematics.
The results are compared with predictions
of chiral perturbation theory where available. The low-energy
expansion of the spin-dependent Compton scattering amplitude 
yields the Gerasimov-Drell-Hearn (GDH) sum rule and relations for the forward spin
polarizabilities (FSPs) of the proton. Our evaluation provides an
empirical verification of the GDH sum rule for the proton, 
and yields empirical values of the proton FSPs. 
For the GDH integral, we obtain $204.5(21.4)$ $\upmu$b, in
agreement with the sum rule prediction: $204.784481(4)$
$\upmu$b. For the FSPs, we obtain:
$\gamma_0=-92.9(10.5) \times 10^{-6}$ fm$^4$,
and  $\bar{\gamma_0}=48.4(8.2) \times 10^{-6}$ fm$^6$,
improving on the accuracy of previous evaluations.
\end{abstract}
\pacs{13.60.Fz - Elastic and Compton scattering,
14.20.Dh - Protons and neutrons,
25.20.Dc - Photon absorption and scattering,
11.55.Hx Sum rules.}
\date{\today}
\maketitle


\section{Introduction}
The forward Compton scattering (CS) off a spin-1/2 target, such as the nucleon, is described
by two complex functions of the photon energy $\nu$:  the spin-independent amplitude, $f(\nu)$,
and  the spin-dependent amplitude, $g(\nu)$. The general requirements of unitarity and causality 
allow one to express these amplitudes in terms of integrals of total photoabsorption cross
sections \cite{GellMann:1954db} (see Refs.\ \cite{Drechsel:2002ar,Hagelstein:2015egb} for reviews). 
At least for the proton, these cross sections are fairly well known by now, and, whereas 
in the previous paper \cite{Gryniuk:2015eza} we obtained the {\it spin-independent} amplitude, here 
we evaluate the {\it spin-dependent} one.
Having both of them, we can reconstruct all observables of the forward CS off the proton.

This is essentially the only way to access the forward CS observables empirically --- a direct observation of strictly forward CS is not possible in practice. Indirectly, the
forward CS can be measured through dilepton photoproduction, where the timelike CS enters prominently in certain kinematics, while the photon virtuality is  small (quasireal CS). However, the only experiment of this kind was done at DESY in 1973 \cite{Alvensleben}, measuring the spin-independent amplitude $f$
at 2.2 GeV.  

The spin-dependent amplitude $g(\nu)$ has not yet been measured through the dilepton photoproduction and 
not much is known about it empirically. 
Until now, only its low-energy expansion has been studied.
The leading-order term yields the Gerasimov--Drell--Hearn (GDH) sum rule \cite{Gerasimov:1965et, Drell:1966jv}, which has
recently been verified for the proton by the GDH Collaboration
\cite{Ahrens:2000bc, Ahrens:2001qt, Dutz:2003mm, Dutz:2004zz}.
The forward spin polarizability (FSP) sum rules, arising at the next two orders, have been evaluated by Pasquini {\it et~al.}~\cite{Pasquini:2010zr}.
In this work, having evaluated 
$g(\nu)$ over a broad energy range, we, too,
consider its low-energy expansion and hence reevaluate
the sum rules. The results are compared
with the previous evaluations in Table~\ref{sumruletest}.

The results for the energy dependence of the amplitude $g(\nu)$ and the observables can be compared with theoretical calculations.
At low the energies we shall compare to the calculations based on chiral perturbation theory ($\chi$PT). More specifically, in the lower
panels  of Figs.~\ref{fig:g}, \ref{fig:ds}, and \ref{fig:2refg}, our results
are compared to the calculations of Lensky {\it et~al.}~\cite{Lensky:2008re,Lensky:2009uv,Lensky:2015awa} done in the manifestly covariant baryon $\chi$PT (B$\chi$PT).  
Other state-of-art $\chi$PT calculations of proton CS, based on the heavy-baryon expansion (HB$\chi$PT) \cite{Griesshammer:2012we,McGovern:2012ew}, were shown \cite{Lensky:2012ag} to be in agreement with
the aforementioned B$\chi$PT calculation, within the estimated theoretical uncertainties. Therefore, in the figure, we only plot one of the 
two, whereas the forward spin polarizabilities can
be compared to either of them in Table~\ref{sumruletest}.

An interesting issue arises when studying the elastic contribution to the sum rules in 
perturbative QED: the low-energy expansion, which goes into the 
sum rule derivation, is inapplicable. It is nonetheless possible
to write down the elastic sum rules for
both the spin-independent~\cite{Gryniuk:2015eza} and the 
spin-dependent cases as covered in the~Appendix.



This paper is organized as follows. In \Secref{overview} we give a brief overview
of the Kramers-Kronig relations for the forward CS amplitudes and the sum rules involving
the spin-dependent amplitude $g$.
In \Secref{fitting} we describe the fitting procedure for the
helicity-difference photoabsorption cross section.
We then evaluate the sum rule integrals and the energy dependence of the spin-dependent amplitudes in \Secref{sumrules},
and present results for the forward CS observables in \Secref{observables}.
Conclusions are given in \Secref{summary}.
In the Appendix we discuss the elastic contribution
to the sum rules in spinor QED at one-loop level.

\section{Forward Compton amplitude and sum rules}
\label{sec:overview}

The Lorentz structure of the {\it forward} CS amplitude for a spin-1/2 target, such as the nucleon, can be decomposed into two terms,\footnote{The tensors can in principle be written
in an explicitly current-conserving form (i.e., such that $q_\mu T^{\mu\nu} =0$); however, additional terms vanish when contracted
with the photon polarization vector $\veps^\mu$, because of $q\cdot \veps=0$.}
\beq
\label{covT}
T^{\mu\nu}(p,q) = - \left[ \mathrm{g}^{\mu\nu} f(\nu)  + \ga^{\mu\nu} g(\nu)
\right], 
\eeq
where g$^{\mu\nu}=\mathrm{diag}(1, -1,-1,-1) $ is the metric tensor and $\ga^{\mu\nu}=\frac{1}{2}[\gamma^\mu,\gamma^\nu]$
is the antisymmetrized product of Dirac matrices. 
The amplitudes $f$ and $g$ are complex functions of the
variable $\nu= p\cdot q/M$, with $p$ and $q$ being the target and 
photon 4-momentum; $p^2=M^2$, $q^2=0$. 
The corresponding helicity amplitudes are found as
\bea
\label{F}
T_{\la_\ga '\la_N ' \la_\ga \la_N } &=&  \chi^{\dagger}_{\la_N'} \left\{
f(\nu) \, \beps_{\la_\ga'}^{*} \cdot \beps_{\la_\ga} \, \right.\\
&&\quad\left.+ \, g(\nu) \,i \, (\beps_{\la_\ga'}^{*} \! \times \beps_{\la_\ga} )\cdot\bsigma
\right\} \chi_{\la_N},\nn
\eea 
where $\la_\ga(\la_\ga')$ is the incoming (outgoing) photon helicity, $\la_N (\la_N')$ is the incoming (outgoing) nucleon helicity, $\beps$ are the photon polarization vectors, and the Pauli spinors $\chi$ denote the target polarization.
In terms of the total helicity $\La=\la_\ga - \la_N $, the
amplitude is diagonal: $T_{\La'\La} = \de_{\La'\La} \, \mathcal{T}_{\La}$.
Hence, there are only two independent helicity amplitudes:
$\mathcal{T}_{\pm 3/2} = (f - g)$ and $\mathcal{T}_{\pm 1/2} = (f + g)$.
The optical theorem (unitarity)  relates the imaginary
part of these amplitudes to the corresponding photoabsorption cross sections,
\beq
\im \mathcal{T}_\La(\nu) = \frac{\nu}{4\pi} \, \sigma_\La(\nu),
\eeq
with $\La= 1/2, 3/2$.
Rewriting these relations for $f$ and $g$, one obtains the
usual,
\begin{subequations}
\label{OptT}
\begin{eqnarray}
\im f(\nu) &=& \frac{\nu}{8\pi} \left[\sigma_{1/2}(\nu)+\sigma_{3/2}(\nu)\right]
\equiv \frac{\nu}{4\pi}\,\si(\nu) , \\
\im g(\nu) &=& \frac{\nu}{8\pi}
\left[\sigma_{1/2}(\nu)-\sigma_{3/2}(\nu)\right]
\equiv -\frac{\nu}{8\pi}\,\De\si(\nu), \quad
\eqlab{unitarity}
\end{eqnarray}
\end{subequations}
where by $\si$ and $\De\si$ we denote, respectively, the unpolarized and the
helicity-difference cross sections of total photoabsorption.
 
Furthermore, the analytic and crossing properties of the
amplitudes $f$ and $g$,  in conjunction with Eq.~(\ref{OptT}), allow one
to write the following dispersion relations,
\begin{subequations}
\begin{eqnarray}
\re f(\nu) &=&-\,\frac{Z^2\alpha}{M} \,+\, \frac{\nu^2}{2\pi^2}\fint_0^\infty\!\dd \nu' \frac{\si(\nu')}{\nu^{\prime \, 2}-\nu^2},\label{KKf}\\
\re g(\nu) &=& -\frac{\nu}{4\pi^2}\fint_0^\infty\! \dd \nu'\,\frac{\nu' \De\si(\nu')}{\nu^{\prime \, 2}-\nu^2}, \label{KK}
\end{eqnarray}
\end{subequations}
where $M$ is the nucleon mass and $Z=1$ or 0 for the proton or neutron, respectively; 
the slashed integral denotes the principal-value integration.
The dispersion relation for the spin-independent amplitude, $f$, 
was discussed in the previous paper \cite{Gryniuk:2015eza}.
Here, we focus on the spin-dependent amplitude, $g$,
and the helicity-difference cross section, $\De\si$, for the proton.

We note that the cross sections
include the CS process itself, as well as multiphoton production.
We neglect those processes since 
they are of course suppressed by at least one order of $\al$
with respect to hadron-production processes, such as the pion photoproduction (more details on leading radiative corrections 
can be found in the Appendix).
The cross sections which exclude electromagnetic radiation will be denoted
by $\si_{\mathrm{abs}}$ and are assumed to begin with the lowest hadron-production threshold. 
We thus deal with the following relation for the spin-dependent forward CS
amplitude,
\bea
\re g(\nu) & = & 
-\frac{\nu}{4\pi^2}\fint_{\nu_0}^\infty\!\dd\nu'\,
\frac{\nu'\De\si_\mathrm{abs}(\nu')}{\nu^{\prime \, 2}-\nu^2},\label{KKg}
\eea
where $\nu_0$ is the pion photoproduction threshold. Its low-energy expansion
yields the following sum rules. 

At the first order,
one obtains the GDH sum rule,
\beq
\label{IGDH}
I_{\mathrm{GDH}} \equiv \int_{\nu_0}^\infty\!\dd\nu\,\frac{\De\si_\mathrm{abs}(\nu)}{\nu}
= \frac{2\pi^2\alpha}{M^2}\kappa^2,
\eeq
where $\kappa$ is the anomalous magnetic moment of the nucleon. Substituting the proton mass $M_p$ and the anomalous magnetic
moment $\kappa_p$ into the right-hand side, we obtain the GDH sum rule prediction
quoted in the last row of Table~\ref{sumruletest}.
 
At the next two orders, one finds the FSP sum rules:
\begin{gather}
\gamma_0 = -\frac{1}{4\pi^2} \int_{\nu_0}^\infty\!\dd\nu\,\frac{\De\si_\mathrm{abs}(\nu)}{\nu^3}, \\
\bar{\gamma_0} = -\frac{1}{4\pi^2} \int_{\nu_0}^\infty\!\dd\nu\,\frac{\De\si_\mathrm{abs}(\nu)}{\nu^5}\,.
\end{gather}

In what follows, we assess the empirical helicity-difference cross section
and evaluate $g(\nu)$, $I_{\mathrm{GDH}}$, $\ga_0$, $\bar\ga_0$ from the
above integrals.

\section{Fit of the polarized photoabsorption cross section}
\label{sec:fitting}

We begin with performing a smooth fit of the 
experimental helicity-difference cross section of total photoabsorption on the proton. The fitting procedure is similar to the one applied for the unpolarized
photoabsorption cross section $\si_{\mathrm{abs}}$ \cite{Gryniuk:2015eza}.
The integration domain is divided into three regions:
\begin{itemize}
 \item[(i)]  {\it low energy},  $\nu \in [\nu_0   \,,\;  \nu_1  ) $;
 \item[(ii)]  {\it medium energy},     $\nu \in [\nu_1   \,,\;  2\;\mathrm{GeV}) $;
 \item[(iii)] {\it high energy}, $\nu \in [2\;\mathrm{GeV} \,,\;  \infty ) $;
\end{itemize}
where $\nu_0$ ($\approx 0.145$ GeV) and $\nu_1$ ($\approx 0.309$ GeV) are thresholds for the single- and double-pion photoproduction,
respectively. A smooth transition between the regions
is implied. 

\begin{figure*}[hbt]
\includegraphics[width=0.74\textwidth]{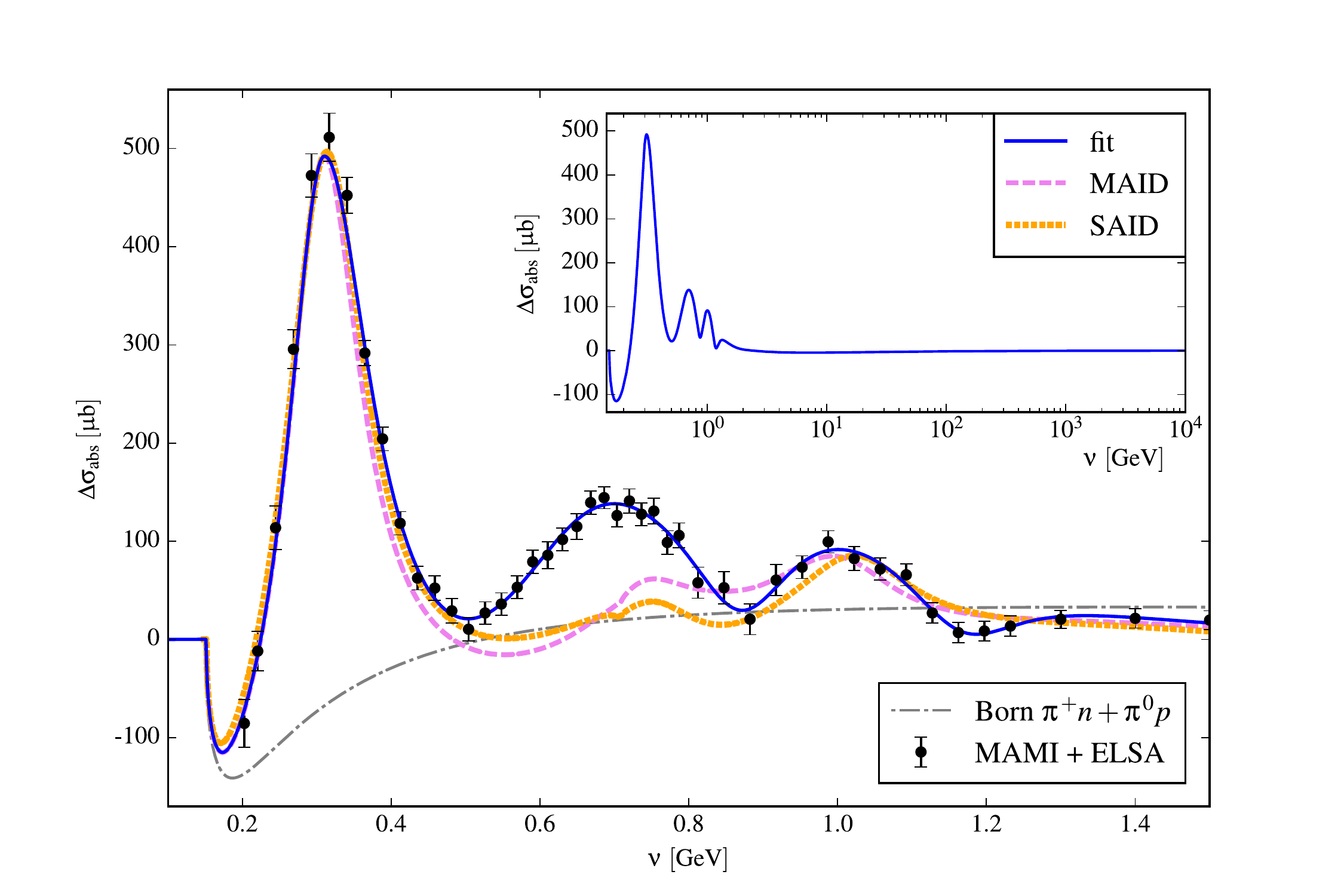}
\caption{
Fit of experimental data for the helicity-difference
cross section of total photoproduction  on the proton.
The solid curve shows our fit. The other curves, according to the
legend, 
show the Born contribution (single-pion production on a pointlike proton), as well as
the results of {\sc maid} \cite{MAID} and {\sc said} \cite{SAID} multipole analyses. 
}
\label{fig:cs_pol}
\end{figure*}

In the {\it low-energy} region, we use the cross sections 
generated by {\sc maid} \cite{MAID} (single-pion production
only). Unfortunately, the {\sc maid} analysis does not provide
any indication of its uncertainty.
In our error estimate, we judiciously apply a 2\% uncertainty to the
{\sc maid} values.

In the {\it medium-energy} region, a fit to the data from the MAMI (Mainz) and ELSA (Bonn) experiments
of the GDH and A2 collaborations \cite{Ahrens:2000bc, Ahrens:2001qt, Dutz:2003mm, Dutz:2004zz}
is applied in the form of a sum of six nonrelativistic Breit-Wigner resonances,
\beq
\label{res_fit}
\De\si_{\mathrm{res}}(W)=\sum_{i=1}^6 A_i \, \frac{\quarter \Ga_i^2}{(W-M_i)^2 + \quarter\Ga_i^2},
\eeq
where $W=\sqrt{s}$ is the invariant mass of the $\gamma p$ system.
Widths ($\Ga$), masses ($M$), and couplings ($A$) are 
treated as free fitting parameters.
The resulting values are given in Table \ref{table:res_par_2}.

\begin{table}[hbt]
\caption{Fitted resonances parameters entering Eq.~(\ref{res_fit}).}
\label{table:res_par_2}
{
\begin{tabular}{c|c|c|c|}
\hline
\,$i$\, & $M_i$ (MeV) & $\Ga_i$ (MeV) & \,$A_i\cdot\quarter\Ga_i^2$\;(nb$\cdot$GeV$^2$)\,\\
\hline
\hline
$1$ & $1210.2$ & $119.3$ & $1047.3 $\\
$2$ & $1405.0$ & $493.5$ & $-9008.4$\\
$3$ & $1460.8$ & $239.8$ & $1964.0 $\\
$4$ & $1585.5$ & $111.7$ & $-226.9 $\\
$5$ & $1616.4$ & $360.7$ & $3829.3 $\\
$6$ & $1752.5$ & $105.0$ & $-103.4 $\\
\hline
\end{tabular}
}
\end{table}

In the {\it high-energy} region, a function of the following Regge form is used:
\beq
\label{regge_fit}
\De\si_\mathrm{Regge}(W)=C_1 \, W^{p_1} + C_2 \, W^{p_2}.\\
\eeq
For $W$ in GeV and the cross section in $\upmu$b,
we use the following fixed parameters \cite{Helbing:2006zp}:
\begin{gather*}
C_1 = -17.05 \pm 2.85 \,, \quad C_2 = 104.7 \pm 14.5 \,, \\
p_1 = -1.16 \pm 0.46 \,, \quad p_2 = -3.32 \pm 0.44 \,.
\end{gather*}

The cross section fitting and the sum rule evaluations are accomplished with the help of the {\sc SciPy} package for {\sc Python}.
We used the weighted nonlinear least-squares optimization procedure of 
{\sc SciPy}'s wrapper around {\sc minpack}'s {\sc lmdif} and {\sc lmder} algorithms.
The latter implement the modified Levenberg-Marquardt algorithm \cite{Levenberg,Marquardt}.

The resulting fit of the helicity-difference photoabsorption cross section is 
shown in Fig.~\ref{fig:cs_pol}.
Also shown is the Born contribution 
for the $\pi^+n + \pi^0p$ photoproduction off a pointlike
proton (with the vanishing anomalous magnetic moment), as well as 
the results of 
{\sc maid} \cite{MAID} and {\sc said} \cite{SAID} multipole analyses.

\section{Evaluation of the integrals}
\label{sec:sumrules}

Having obtained the fit of the total photoproduction cross section, we proceed to the evaluation of the GDH and FSP sum rules, 
and ultimately of the forward CS amplitude.
Our results for the sum rule integrals are presented in Table \ref{sumruletest}, where they can be compared with some of the previous empirical evaluations, as well as the
recent $\chi$PT results. 

To estimate the uncertainty of our fits and dispersive integrals, we
compute the covariance matrix of the fitted parameters.
In the medium-energy region, the covariance matrix 
is simply obtained based on the experimental uncertainties of the data 
points. In the high-energy region, we make use of the uncertainties for the Regge parameters from Ref.\ \cite{Helbing:2006zp},
assuming that these four parameters are uncorrelated.
We then apply the standard, linear error propagation to find
the uncertainty of the dispersive integrals.

In the low-energy region, where the cross sections are not fit
but obtained from the partial-wave analyses, we 
judiciously estimate the systematic error of each of the photoabsorption cross sections to be 2\% of the magnitude of the unpolarized cross section, $\si(\nu)$. As the result, the error on $\De\si(\nu)$ is equal to 4\% of $\si(\nu)$. This error is then	linearly propagated to the dispersion integrals.

Within the calculated uncertainties, our evaluation appears to be consistent with the previous ones, as well as with the GDH sum rule value quoted in the bottom part of Table \ref{sumruletest}. The discrepancy in the central value
of the GDH integral can be traced back to the fact that 
our fit at the $\De$-resonance peak happens to lie well below the 
central value of the data point; see Fig.~\ref{fig:cs_pol}. 
In particular, the GDH
Collaboration \cite{Dutz:2004zz} obtains $(254 \pm 5 \pm 12)$ $\upmu$b in the interval of available data 
(i.e., $0.2<\nu <2.9$ GeV), while our fit of the same data
yields $(246.4 \pm 6.8)\,\upmu$b.

\begin{table}[tbh]
\begin{ruledtabular}
\centering 
\caption{Empirical evaluations of the GDH and FSP integrals.}
    \begin{tabular}{|c|c|c|c|}
    & $I_\mathrm{GDH}$ & $\gamma_0$ & $\bar{\gamma_0}$ \\
   	& ($\upmu$b) & ($10^{-6}$ fm$^4$) & ($10^{-6}$ fm$^6$) \\
	\hline
    GDH \& A2 \cite{Ahrens:2001qt, Dutz:2004zz} & $\approx 212$ & $\approx -86$ & \\
    Helbing \cite{Helbing:2006zp} & $212 \pm 6 \pm 12$ &&\\
    Bianchi-Thomas \cite{Bianchi:1999qs} & $207 \pm 23$ & & \\
    Pasquini {\it et~al.}~\cite{Pasquini:2010zr} & $210 \pm 6 \pm 14$ & $-90 \pm 8 \pm 11$ & $60 \pm 7 \pm 7$ \\
     This work & $204.5 \pm 21.4$ & $-92.9 \pm 10.5 $ & $ 48.4 \pm 8.2 $ \\
	\hline
     GDH sum rule & $204.784481(4)$\footnote{Right-hand side of Eq.~(\ref{IGDH}) with CODATA \cite{CODATA} values of proton $M$ and $\kappa$.} & & \\
     B$\chi$PT \cite{Lensky:2015awa} &  &$- 90 \pm 140$ & $110\pm 50$  \\
     HB$\chi$PT \cite{McGovern:2012ew} &&$- 260 \pm 190$ &
    \end{tabular}
	\label{sumruletest}
\end{ruledtabular}
\end{table}

\begin{figure}[hbt]
\includegraphics[width=0.5\textwidth]{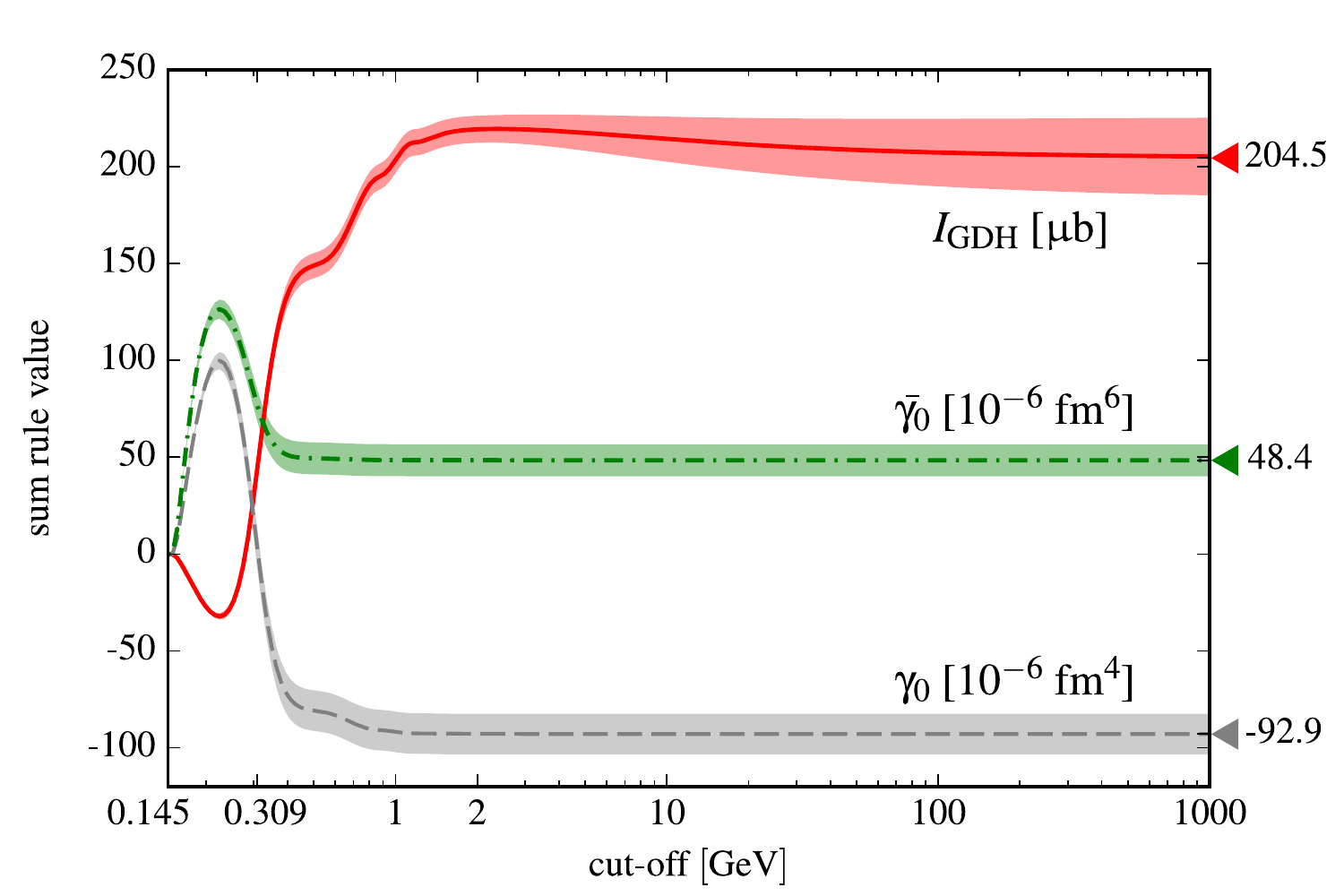}
\caption{
The GDH  and FSP integrals
as a function of the upper integration bound.
Bands represent estimated errors. Asymptotic values of the integrals are displayed
on the right and marked with colored triangles.
}
\label{fig:135}
\end{figure}

Table \ref{sr_regions} 
shows the contributions from each of the three energy regions. 
One can clearly see that the high-energy contribution
is negligible for the FSPs.
A more detailed behavior of the running sum rule integrals
(functions of the cut-off --- the upper integration bound)
can be seen in Fig.~\ref{fig:135}. One can see the
good convergence properties of all the integrals. 
It is interesting to observe the significant cancellations 
between the contribution below and above $0.2$ GeV. 

We note that the main contribution to the estimated uncertainty of the GDH integral
comes from the high-energy Regge behavior,
which is possibly both due to the fact that parameters seem to be not well ``fixed''
and because we have used a simplified covariance matrix estimation for these parameters.
As for the higher-order sum rules, it appears that the main contribution
to the uncertainty comes from our assumption about the systematic uncertainty of the partial-wave analyses (low-energy region).

\begin{table}[tbh]{
\caption{Contributions to the GDH and FSP integrals by regions.}
   \label{sr_regions}
	\centering 
	\begin{tabular}{|c|c|c|c|}
	\hline
\diagbox{Sum Rule}{Region} & {\it low-energy} & {\it medium-energy}  & {\it high-energy}  \\
	\hline
	$I_\mathrm{GDH}$ \; ($\upmu$b) & $43.6 \pm 6.0$ & $175.7 \pm 3.7$ & $-14.8 \pm 19.9$ \\
    $\gamma_0$ \; ($10^{-6}\,$ fm$^4$) & $3.6 \pm 10.3$ & $-96.5 \pm 2.0$ & $(2 \pm 7) \times 10^{-2}$ \\
    $\bar{\gamma_0}$ \; ($10^{-6}\,$ fm$^6$) & $77.1 \pm 8.2$ & $-28.7 \pm 0.6$ & $(2 \pm 36) \times 10^{-5}$ \\
	\hline
	\end{tabular}
	\par} 
\end{table}

We next evaluate the entire spin-dependent amplitude 
$g(\nu)$. In order to improve on the accuracy, 
we use the subtracted dispersion relation:
\beq
\label{KKS}
\re g(\nu) \,=\, -\frac{\alpha \kappa^2}{2M^2}\nu  \,-\, \frac{\nu^3}{4\pi^2}\fint_{\nu_0}^\infty\!\dd\nu'\,\frac{\Delta\si_\mathrm{abs}(\nu')}{(\nu'^{\,2}-\nu^2)\,\nu'}.
\eeq
The only difference with the unsubtracted one, Eq.~(\ref{KKg}), is accuracy. Indeed, the subtraction replaces the 
value of the GDH integral (see ``This work'' in Table \ref{sumruletest}) by the much more accurate GDH sum rule value (next row)
and leads to the smaller uncertainty.

\begin{figure}[hbt]
\includegraphics[width=0.5\textwidth]{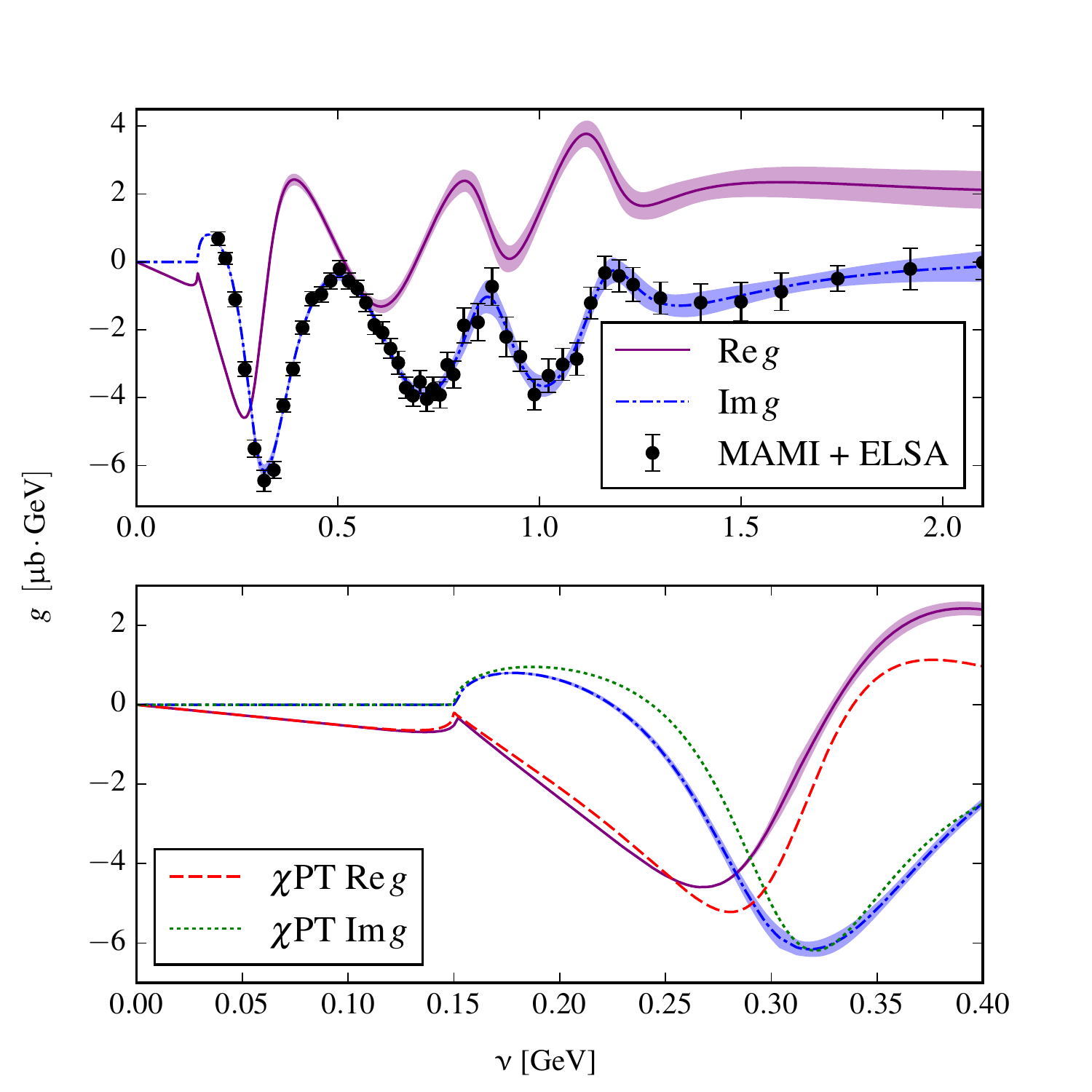}
\caption{
Spin-dependent amplitude $g(\nu)$ obtained from numerical integration of the fit of data for the helicity-difference photoproduction cross section.
Dashed and dotted curves in the bottom panel are the
B$\chi$PT predictions of Ref.~\cite{Lensky:2015awa}.
Bands represent the error estimate.
}
\label{fig:g}
\end{figure}

The remaining integral in Eq.~(\ref{KKS}) converges very fast
in the considered energy range. The resulting amplitude
is plotted in Fig.~\ref{fig:g}. The upper panel shows
the real and imaginary parts in the energy range where 
the data (for the imaginary part) are available.

The lower panel of Fig.~\ref{fig:g} zooms into the lower energy range where
our results can be compared with 
next-next-to-leading order $\chi$PT calculations
of Lensky {\it et~al.}~\cite{Lensky:2015awa}.
One notes here that
the imaginary parts differ appreciably at energies
around 0.25 GeV. Nevertheless,  their integrals (i.e.,
the real parts) agree perfectly at low $\nu$.
This is a ``scientific miracle'' 
of the effective field theory ---
the low-energy quantities are well described, even
though they are obtained as loop or dispersive integrals which
include higher-energy domains where the theory is inapplicable.

\section{Observables}
\label{sec:observables}

Given both amplitudes, $f(\nu)$ and $g(\nu)$, one can reconstruct the energy dependence
of the forward CS observables.
The differential cross section of the forward CS in the laboratory frame is expressed as follows:
\beq
\eqlab{dsi}
\frac{\dd \si}{\dd\Omega_\mathrm{lab}}\bigg|_{\th=0} =
 |f|^2 +|g|^2.
\eeq
The other nonvanishing observables involve the photon and nucleon
spin polarization along the $z$ axis. When only the initial particles
are polarized, the corresponding asymmetry
is known as $\Si_{2z}$ and can be defined as
\beq
\Si_{2z} = \frac{\dd\si_{3/2} - \dd\si_{1/2}}
{\dd\si_{3/2} + \dd\si_{1/2}},
\eeq
where $\si_\La$ denotes the doubly polarized CS cross section with $\La$ being the combined helicity of the initial state. When the initial
photon and the final nucleon are polarized, the asymmetry is called $\Si_{2z'}$.
For the forward scattering, all asymmetries involving such pairwise polarization are equal and in terms of $f$ and $g$ are given as follows:
\beq
\eqlab{Si2z}
\Si_{2z}\big|_{\th=0} = -\frac{2 \re(f g^*)}{|f|^2 +|g|^2}.
\eeq

\begin{figure}[t]
\includegraphics[width=0.5\textwidth]{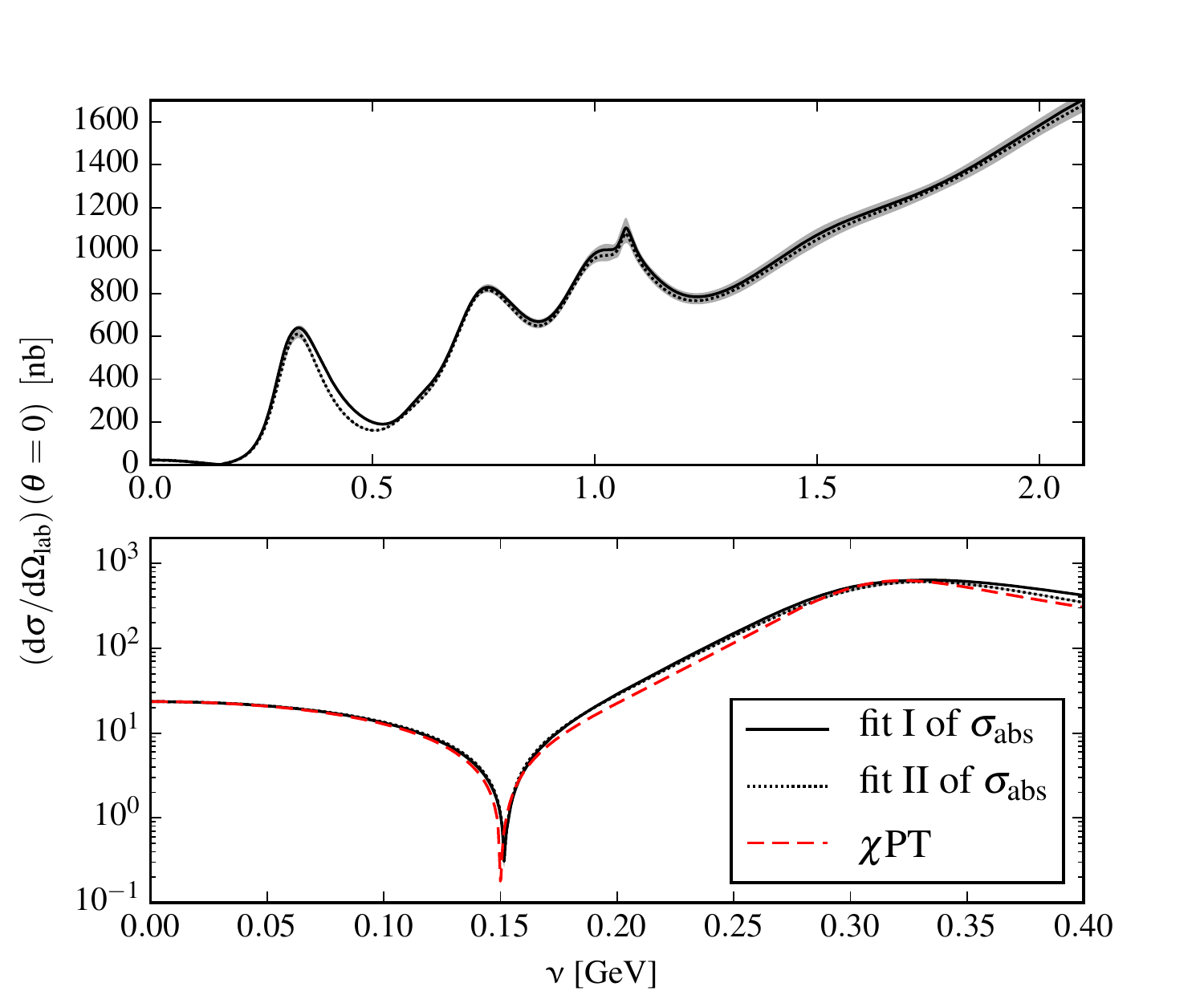}
\caption{
Differential cross section of the forward CS off the proton
for the
two distinctive fits of the unpolarized photoabsorption cross section,
obtained in Ref.~\cite{Gryniuk:2015eza}.
The solid gray band represents the resulting estimated uncertainty on
the observable.
The red dashed curve shows the B$\chi$PT calculation
of Ref.~\cite{Lensky:2015awa}.
}
\label{fig:ds}
\end{figure}

\begin{figure}[t]
\includegraphics[width=0.5\textwidth]{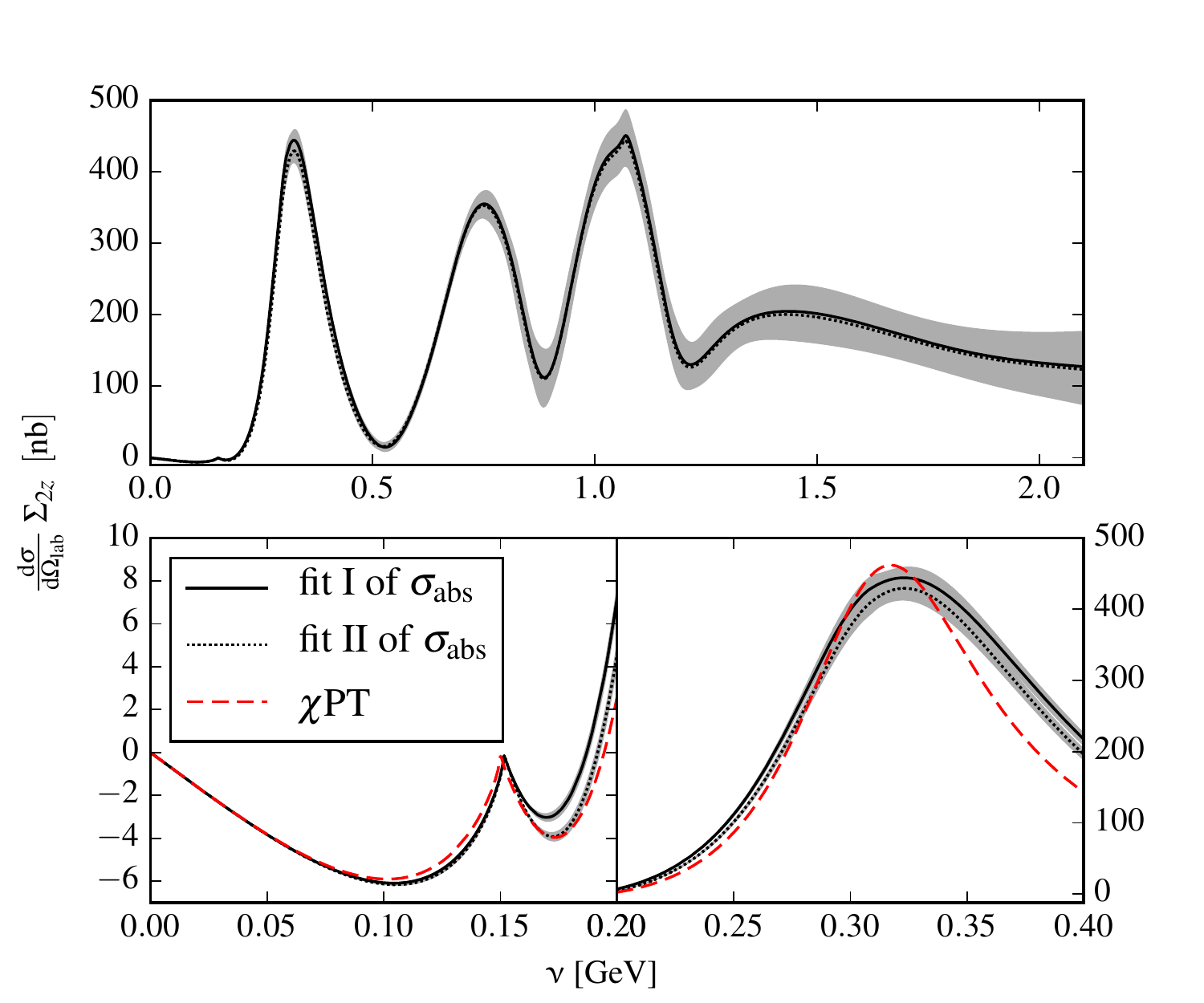}
\caption{Unpolarized differential cross section multiplied with
$\Sigma_{2z}$-asymmetry for the forward CS off the proton.
The two distinctive fits of the unpolarized photoabsorption cross section are obtained in Ref.~\cite{Gryniuk:2015eza}.
The solid gray band represents the resulting estimated uncertainty on
the observable.
The red dashed curve shows the B$\chi$PT calculation
of Ref.~\cite{Lensky:2015awa}.
}
\label{fig:2refg}
\end{figure}

In Table \ref{tab:values_forward_amps}, we list the empirical values of the real part of the forward amplitudes $f(\nu)$ and $g(\nu)$ for several
values of $\nu$; for other values, see Ref.\ \cite{Supplement}. Below the pion-production threshold ($\nu < m_\pi + m_\pi^2/2M_p$), where 
the amplitudes are real, the observables can simply be obtained
by substituting these values in \Eqref{dsi} and \eref{Si2z}.
Above the pion threshold, one obviously needs the photoabsorption 
cross sections [cf.~Eq.~(\ref{OptT})], which we include in the Supplemental Material \cite{Supplement}.

\begin{table}[htp]
\caption{Empirical values of the forward amplitudes
for proton CS in units of $\upmu{\textrm{b}}\cdot\mathrm{GeV}$.
}
\label{tab:values_forward_amps}
\begin{center}
\begin{tabular}{|c|c|c|c|}
\hline
\rule{0pt}{3.5ex} 
$\nu$ (MeV) & $\re f_\mathrm{(fit\,I)}$  & $\re f_\mathrm{(fit\,II)}$  & \;$\re g$\;  \\
\hline
\rule{0pt}{3.ex} 
$50$ & $-2.84 \pm 0.01$ & $-2.85 \pm 0.01$ & $-0.26 \pm 0.01$ \\
\rule{0pt}{3.ex}
$60$ & $-2.76 \pm 0.01$ & $-2.76 \pm 0.01$ & $-0.32 \pm 0.01$ \\
\rule{0pt}{3.ex}
$70$ & $-2.65 \pm 0.01$ & $-2.66 \pm 0.01$ & $-0.37 \pm 0.01$ \\
\rule{0pt}{3.ex}
$79$ & $-2.54 \pm 0.01$ & $-2.56 \pm 0.01$ & $-0.42 \pm 0.01$ \\
\rule{0pt}{3.ex}
$90$ & $-2.38 \pm 0.01$ & $-2.40 \pm 0.01$ & $-0.48 \pm 0.01$ \\
\rule{0pt}{3.ex}
$100$ & $-2.21 \pm 0.01$ & $-2.23 \pm 0.01$ & $-0.54 \pm 0.01$ \\
\rule{0pt}{3.ex}
$110$ & $-2.00 \pm 0.01$ & $-2.03 \pm 0.01$ & $-0.59 \pm 0.01$ \\
\rule{0pt}{3.ex}
$120$ & $-1.75 \pm 0.01$ & $-1.78 \pm 0.01$ & $-0.64 \pm 0.01$ \\
\rule{0pt}{3.ex}
$130$ & $-1.45 \pm 0.02$ & $-1.48 \pm 0.02$ & $-0.68 \pm 0.01$ \\
\rule{0pt}{3.ex}
$135$ & $-1.27 \pm 0.02$ & $-1.30 \pm 0.02$ & $-0.69 \pm 0.01$ \\
\rule{0pt}{3.ex}
$145$ & $-0.79 \pm 0.03$ & $-0.83 \pm 0.03$ & $-0.65 \pm 0.02$ \\
\rule{0pt}{3.ex}
$150$ & $-0.39 \pm 0.03$ & $-0.42 \pm 0.03$ & $-0.52 \pm 0.02$ \\
\rule{0pt}{3.ex}
$160$ & $0.04 \pm 0.04$ & $0.00 \pm 0.04$ & $-0.68 \pm 0.03$ \\
\rule{0pt}{3.ex}
$180$ & $0.48 \pm 0.04$ & $0.34 \pm 0.04$ & $-1.53 \pm 0.03$ \\
\rule{0pt}{3.ex}
$200$ & $1.13 \pm 0.04$ & $0.92 \pm 0.04$ & $-2.35 \pm 0.03$ \\
\rule{0pt}{3.ex}
$225$ & $2.25 \pm 0.05$ & $1.94 \pm 0.05$ & $-3.38 \pm 0.03$ \\
\rule{0pt}{3.ex}
$245$ & $3.29 \pm 0.05$ & $2.87 \pm 0.05$ & $-4.14 \pm 0.03$ \\
\rule{0pt}{3.ex}
$265$ & $4.09 \pm 0.04$ & $3.55 \pm 0.05$ & $-4.59 \pm 0.03$ \\
\rule{0pt}{3.ex}
$275$ & $4.14 \pm 0.03$ & $3.58 \pm 0.04$ & $-4.53 \pm 0.03$ \\
\rule{0pt}{3.ex}
$300$ & $2.15 \pm 0.14$ & $1.82 \pm 0.16$ & $-3.00 \pm 0.18$ \\
\rule{0pt}{3.ex}
$325$ & $-2.00 \pm 0.22$ & $-2.41 \pm 0.24$ & $-0.44 \pm 0.22$ \\
\rule{0pt}{3.ex}
$440$ & $-7.12 \pm 0.16$ & $-6.27 \pm 0.11$ & $1.81 \pm 0.15$ \\
\rule{0pt}{3.ex}
$580$ & $-2.93 \pm 0.18$ & $-2.47 \pm 0.11$ & $-1.18 \pm 0.19$ \\
\rule{0pt}{3.ex}
$750$ & $-7.12 \pm 0.14$ & $-6.63 \pm 0.13$ & $1.45 \pm 0.25$ \\
\rule{0pt}{3.ex}
$1000$ & $-9.73 \pm 0.19$ & $-9.32 \pm 0.19$ & $1.46 \pm 0.36$ \\
\rule{0pt}{3.ex}
$2200$ & $-10.35 \pm 0.21$ & $-9.97 \pm 0.21$ & $2.10 \pm 0.55$ \\
\hline
\end{tabular}
\end{center}
\label{default}
\end{table}%

The resulting unpolarized and helicity-difference
CS differential cross sections at zero angle are shown in Figs.~\ref{fig:ds}
and \ref{fig:2refg}, respectively. The lower panels show the
comparison with $\chi$PT predictions at lower energies. Note the
logarithmic scale in the case of the 
unpolarized cross section,  Fig.~\ref{fig:ds} (lower panel). 
Incidentally, both $f$ and $g$ amplitudes happen to nearly vanish at the pion-production threshold, and so do the cross sections.

The error estimation for the observables is accomplished with the same basic approach
as for the case of dispersion integrals, described above.
We account for the uncertainty contribution of both $f$ and $g$.

Once again, a remarkable agreement with the $\chi$PT calculations at low energies is observed.
For the case of the beam asymmetry, however, the cusp at $\nu\approx 151.5$ MeV,
caused by the second pion production threshold (namely the $\pi^+n$ channel),
appears to be somewhat sharper than the one obtained within $\chi$PT.
This is mainly because the 
neutral and charged pion production thresholds 
coincide in the $\chi$PT calculation due to 
unbroken isospin symmetry.

\section{Conclusion}
\label{sec:summary}

We have obtained a first complete model-independent
evaluation of the forward Compton scattering off the 
proton. Our results are based on dispersive sum rules
for which  the empirical total photoabsorption
cross sections serve as input. 
Putting together the fits of the unpolarized 
photoabsorption cross section obtained earlier 
\cite{Gryniuk:2015eza} and the helicity-difference
cross section obtained here, we have computed
the sum rule integrals as well as the energy dependence 
of the forward CS amplitudes. 

The existing database for the helicity-difference photoabsorption cross section 
($\De \si_{\mathrm{abs}}$) is not as comprehensive as for the  unpolarized cross section ($\si_{\mathrm{abs}}$).
It consists of only the MAMI (Mainz) and ELSA (Bonn) 
experimental data between $0.2$ and $2.9$ GeV. Nevertheless, 
it proved to be sufficient for a reliable 
evaluation of the spin-dependent amplitude, 
and subsequently the observables, for 
photon lab energies up to several GeV. The influence
of the experimentally unknown high-energy behavior of the 
helicity-difference cross section is largely diminished
by using a subtraction in the form of the
GDH sum rule, cf.~Eq.~(\ref{KKS}).

Considering the low-energy expansion of the 
forward spin-dependent amplitude, we have (re)evaluated
the GDH integral and the two first forward spin
polarizabilities. These results are 
in agreement with the previous evaluations, cf.~Table \ref{sumruletest}.

As explained in the Introduction, it is very difficult
to access the forward CS experimentally. Most 
of the CS data for the proton are in fact 
obtained at scattering angles of 90 deg and above.
On the other hand, the relation of the forward 
CS to the photoabsorption
data exploited here is exact, and this sort of evaluation
is the next best thing to the real CS data. 
In some aspects, it is even better (e.g., accuracy and
costs). The main advantage is that we directly access
the amplitudes, rather than observables,
and hence one has, for example, simpler (linear, rather than bilinear) constraints on the multipole amplitudes. The multipole analysis
of the proton CS data using the stringent constraints of the forward
scattering is a subject for near-future work.

\acknowledgements
This work was supported by the Deutsche Forschungsgemeinschaft through the Collaborative Research Center SFB 1044 [The Low-Energy Frontier of the Standard Model] and the Graduate School DFG/GRK 1581
[Symmetry Breaking in Fundamental Interactions].\\

\appendix

\section{Sum rules for elastic contribution in spinor QED}
\label{sec:spinorQEDExample}
\begin{figure}[h]
  \centering
  \includegraphics[width=0.35\columnwidth]{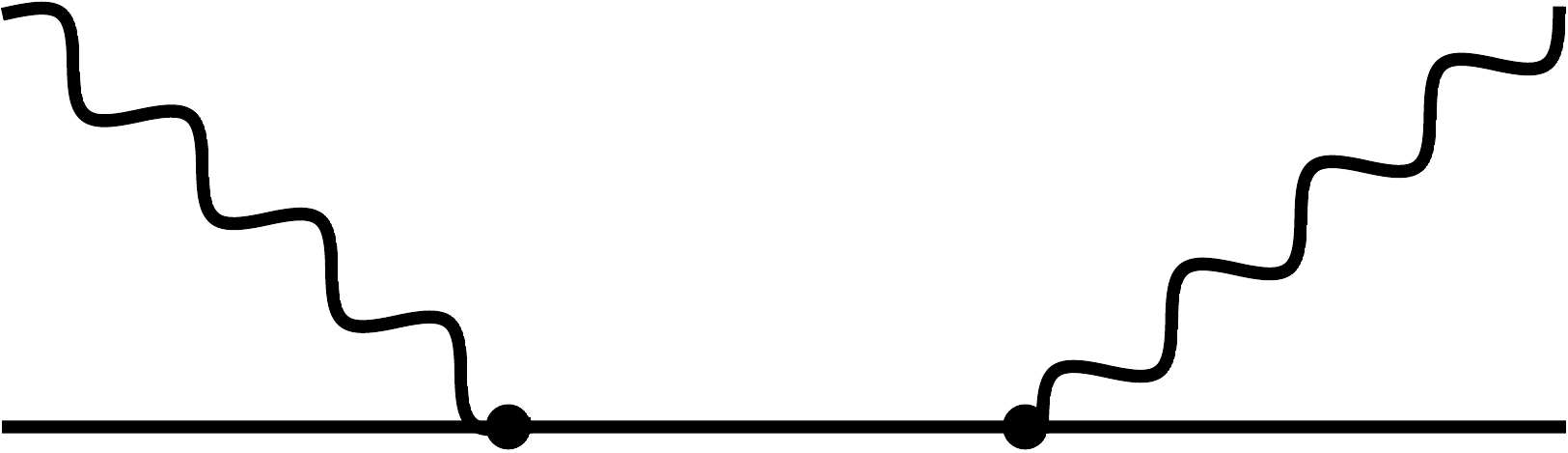} \hspace{5mm}
    \includegraphics[width=0.35\columnwidth]{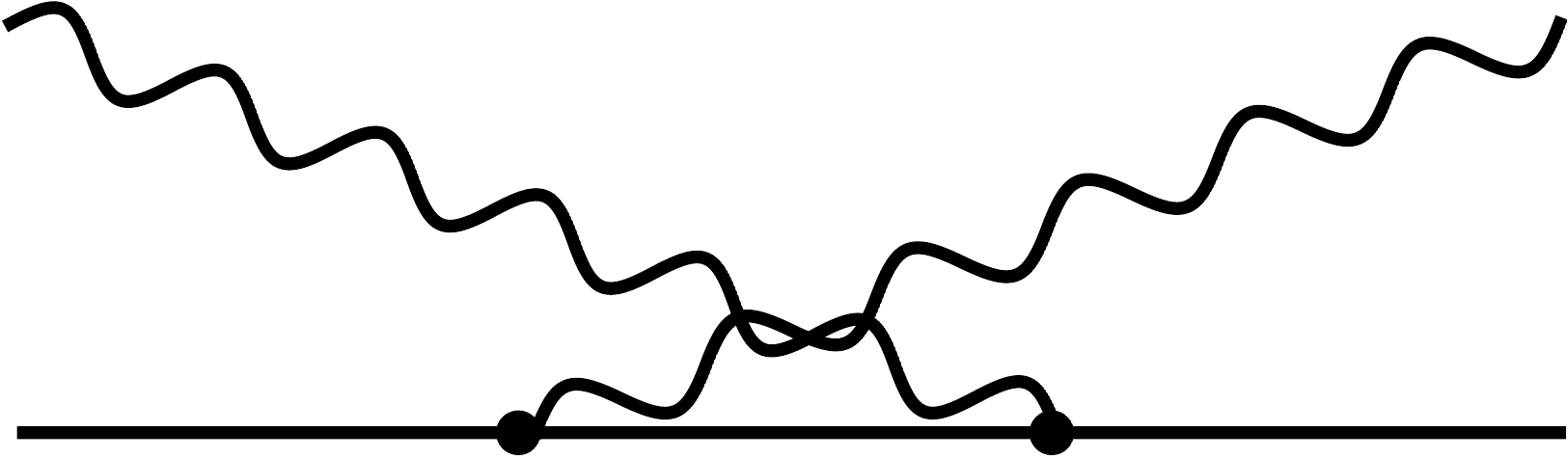}
  	\caption{Tree-level CS graphs.}  
	\figlab{TreeLevelCS}
\end{figure} 
 
\begin{figure}[h]
  \centering
    \includegraphics[width=0.35\columnwidth]{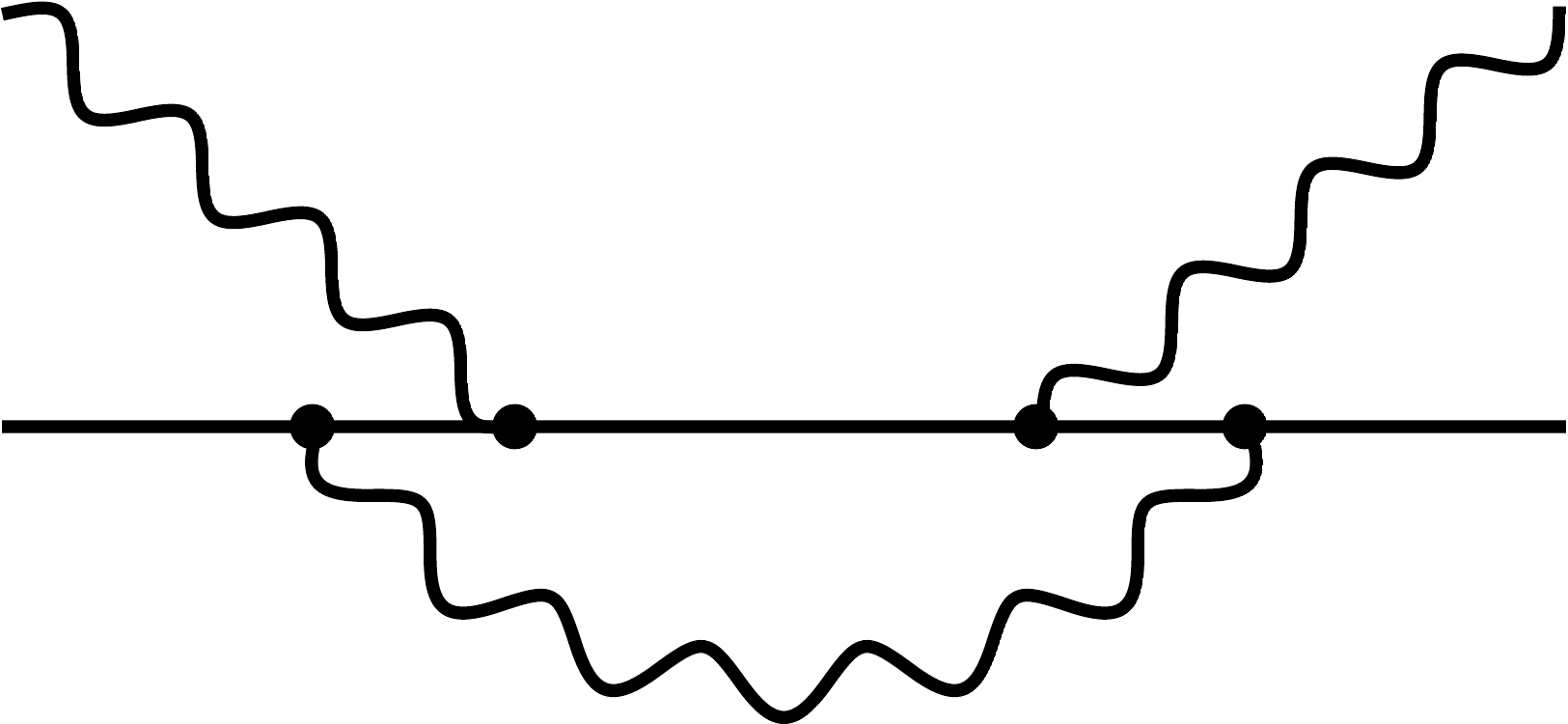}\hspace{5mm}
      \includegraphics[width=0.35\columnwidth]{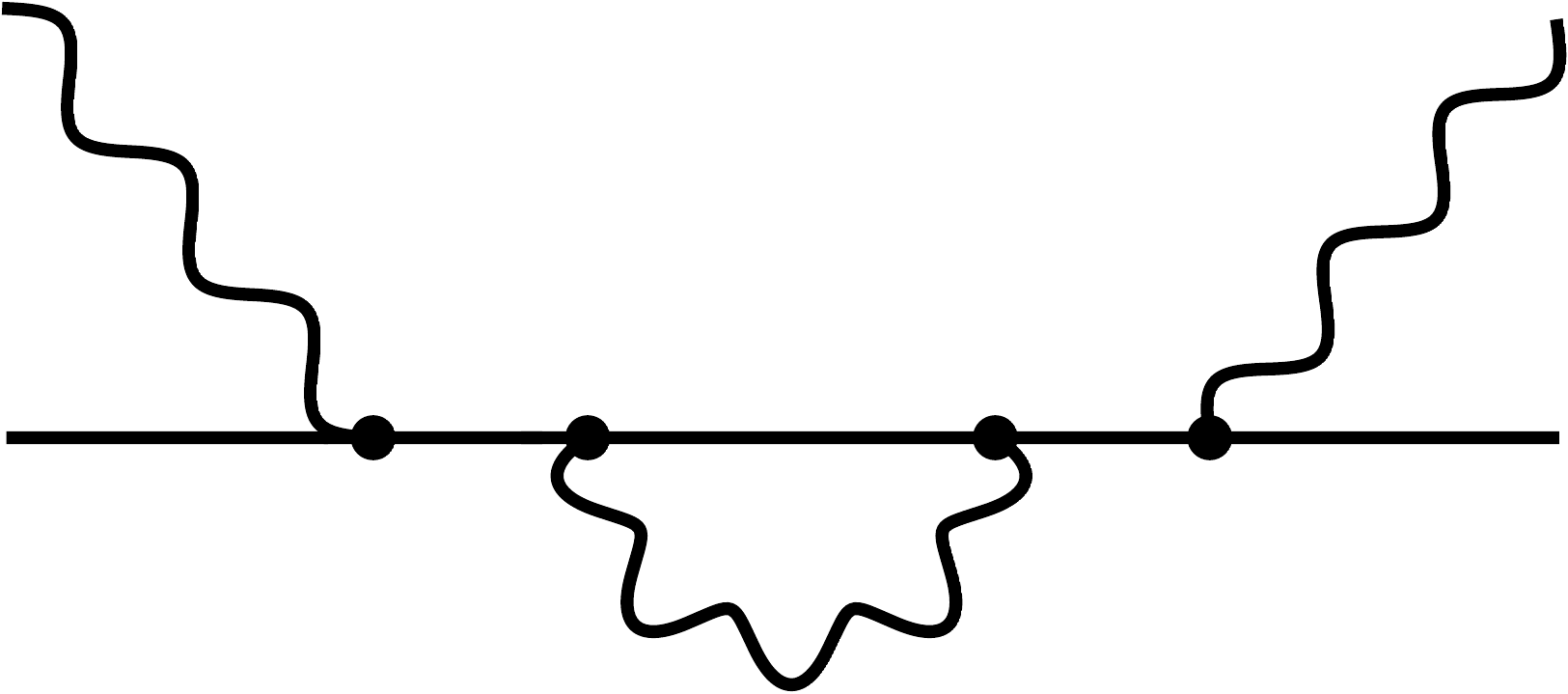}\\[1mm]          \includegraphics[width=0.35\columnwidth]{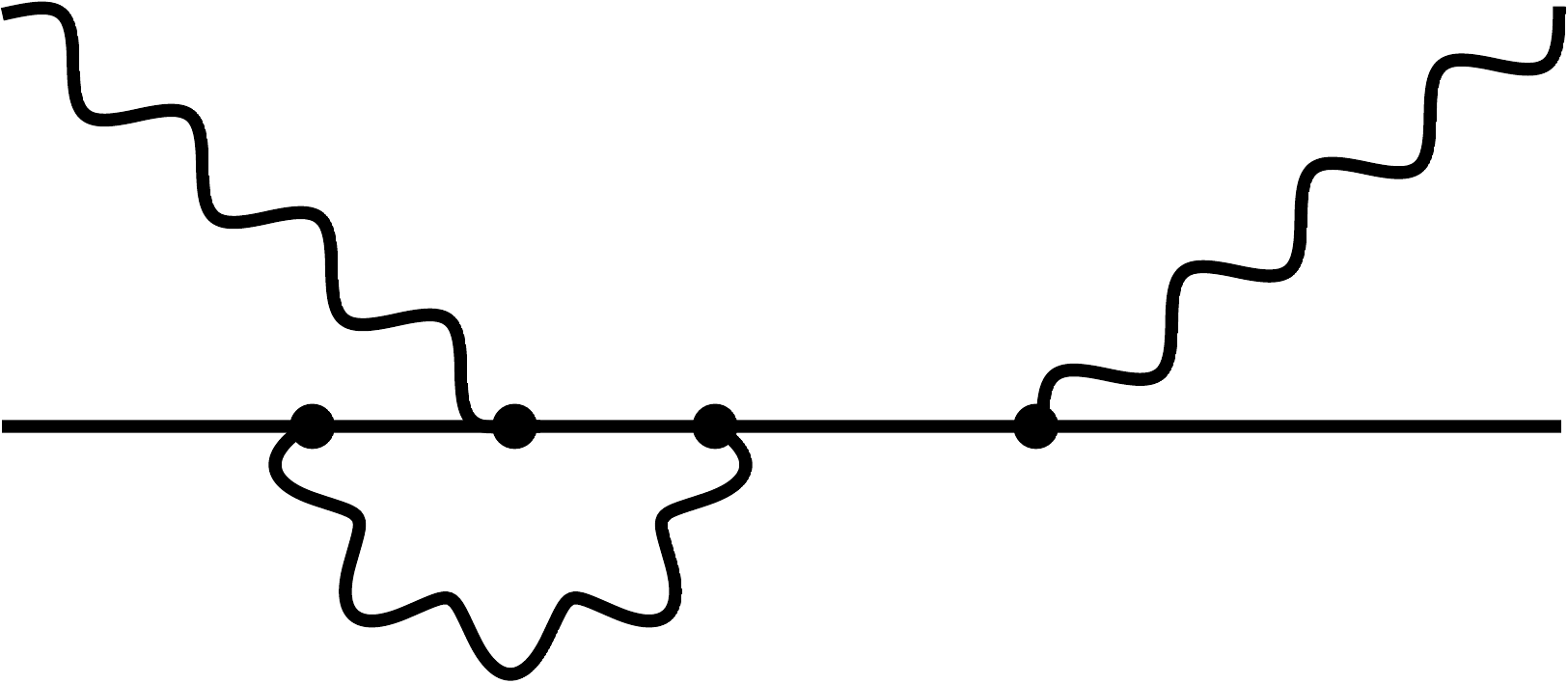}\hspace{5mm}
      \includegraphics[width=0.35\columnwidth]{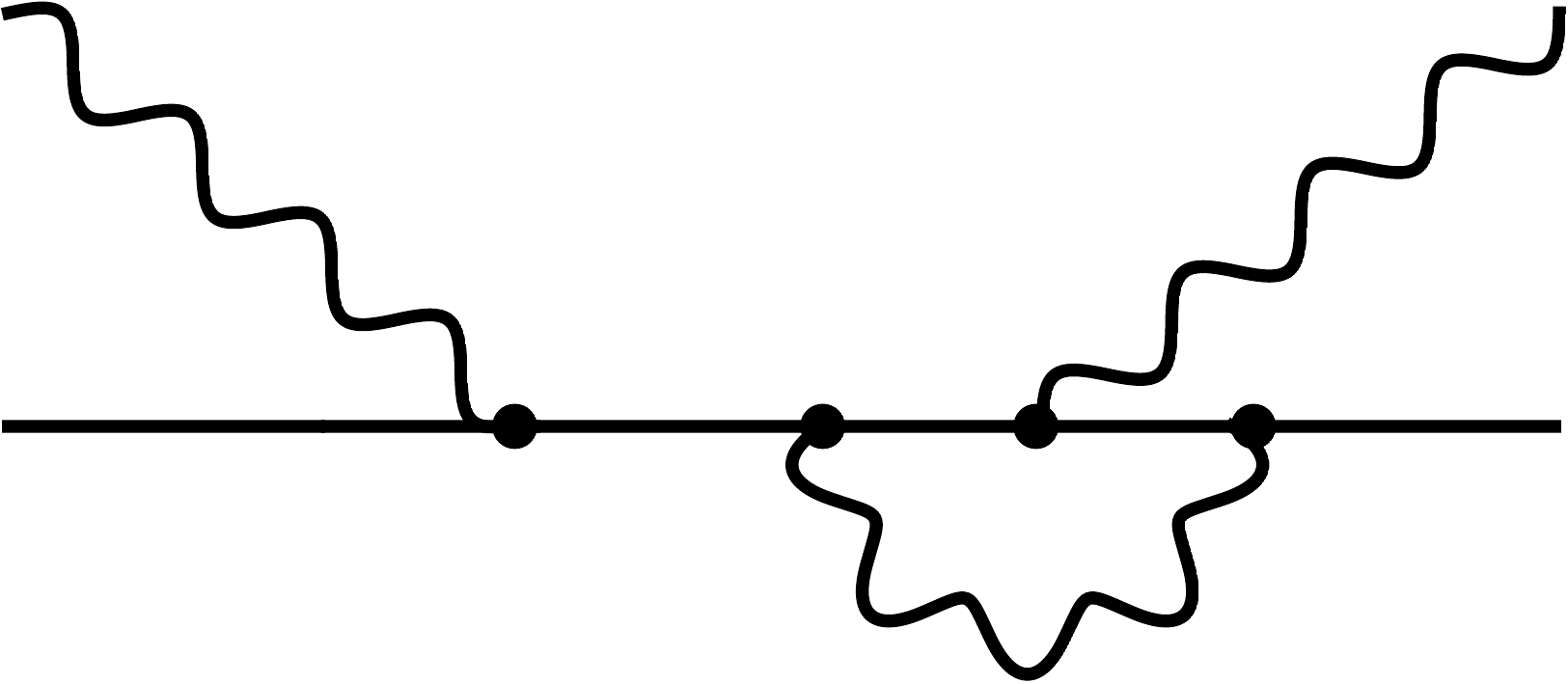}\\[1mm]
 	\caption{One-loop graphs contributing to the forward CS. Graphs obtained from these by crossing of the photon lines are included, too.}  
	\label{fig:IMdiagrams}
\end{figure}
Let us examine the CS (elastic) contribution to the sum rules in spinor QED at $O(\al^2)$. Consider the scattering of a photon from a charged spin-1/2 particle with mass $M$. The helicity amplitudes are expressed in terms of the Feynman amplitude as
\begin{equation}
\eqlab{helamp}
T_{\la_\ga'\la_N'\la_\ga \la_N} 
= \bar u_{\la_N'} (\boldsymbol{p}') \, \veps^{ \ast}_{\la_\ga'}(q')\cdot T(q',q,p',p)\cdot \veps_{\la_\ga}(q)
\, u_{\la_N} (\boldsymbol{p})\,,\nn
\end{equation}
where we denote the helicity and 4-momentum of the incoming (outgoing) photon by $\la_\ga(\la_\ga')$ and $q(q')$ and the helicity and momentum of the incoming (outgoing) spin-1/2 particle by $\la_N (\la_N')$ and $p(p')$. The spinors are normalized according to
\bea 
&& \bar u_{\la_N'}( \boldsymbol{p} )\, u_{\la_N} (\boldsymbol{p})
= \delta_{\la_N' \la_N},
\eea
where $\boldsymbol{p}$ is the nucleon 3-momentum. In general, there are six independent helicity amplitudes for the CS process. The $O(\al^2)$ unpolarized and double-polarized cross sections \cite{Holstein:2005db} can be deduced from the tree-level helicity amplitudes \cite{Tsai:1972sg,Milton:1972sh},
\begin{widetext}
\begin{subequations}
\eqlab{crossSectionTree}
\bea
\sigma^{(2)}&=&\frac{2\pi\al^2}{M^2}\left\{\frac{1+x}{x^3}\left[\frac{2x(1+x)}{1+2x}-\ln(1+2x)\right]+\frac{1}{2x}\ln(1+2x)-\frac{1+3x}{(1+2x)^2}\right\},\\
\Delta \sigma^{(2)}&=&-\frac{2\pi\al^2}{M^2x}\left\{\left[1+\frac{1}{x}\right]\ln(1+2x)-2\left[1+\frac{x^2}{(1+2x)^2}\right]\right\},
\eea
\end{subequations}
with $x=\nu/M$. We note that in the low-energy limit the total unpolarized cross section, $\si$, reproduces the Thomson cross section,
 $ \si^{(2)}(0)  = 8\pi \al^2/3M^2$,
a result that is unaltered by loop corrections. In the same limit, the helicity-difference cross section, $\Delta \si$, is vanishing.

In the forward limit, only two helicity amplitudes are nonvanishing. These are the amplitudes without spin flip: $T_{+1\,+\nicefrac{1}{2}\,+1\,+\nicefrac{1}{2}}$ and $T_{-1\,+\nicefrac{1}{2}\,-1\,+\nicefrac{1}{2}}$. They can be used to reconstruct the spin-dependent and spin-independent forward CS amplitudes: 
\begin{subequations}
\bea
f&=&\frac{1}{2}\left[T_{+1\,+\nicefrac{1}{2}\,+1\,+\nicefrac{1}{2}} +T_{-1\,+\nicefrac{1}{2}\,-1\,+\nicefrac{1}{2}}\right],\\
g&=&\frac{1}{2}\left[T_{+1\,+\nicefrac{1}{2}\,+1\,+\nicefrac{1}{2}}-T_{-1\,+\nicefrac{1}{2}\,-1\,+\nicefrac{1}{2}}\right].
\eea
\end{subequations}
At tree level
(\Figref{TreeLevelCS}), the spinor QED calculation yields $f^{(1)}(\nu) = -\al/M$ and $g^{(1)}(\nu) =0$, where the superscript indicates the order of $\al$. Next, we consider the one-loop corrections (\Figref{IMdiagrams}).
Tsai {\it et~al.}~\cite{Tsai:1972sg,Milton:1972sh} calculated the Compton scattering helicity amplitudes up to $O(\al^2)$. From their result, we obtain the forward CS amplitudes:
\begin{subequations}
\bea
f^{(2)}(x)&=&\frac{\al^2}{4\pi M}\left\{\frac{24 x ^2\left(1-3 x ^2\right)+\pi ^2 \left(4 x ^4+8 x ^3-9 x ^2-2 x +2\right)}{6 x ^2 \left(1-4 x ^2\right)}-\frac{4 x ^2 \left(4 x ^2-3\right)}{\left(4 x ^2-1\right)^2}\,\ln 2x\right.\\
&&\left.-\frac{x ^2-2 x -2}{x ^2}\left[ \ln 2 x \ln(1+2x)+\mathrm{Li}_2\left(-2x\right)\right]+\frac{x ^2+2 x -2}{x ^2}\,\mathrm{Li}_2\left(1-2x\right)\right\}+\frac{i M x}{4\pi}\, \sigma^{(2)}(x),\nn\\
g^{(2)}(x)&=&\frac{\al^2}{4\pi M}\left\{\frac{12 x ^2+\pi ^2 \left(4 x ^3-4 x ^2-x +1\right)}{6 x  \left(4 x ^2-1\right)}-\frac{16 x ^3}{\left(4 x ^2-1\right)^2}\,\ln 2x\right.\\
&&\left.-\frac{x +1}{x }\left[ \ln 2 x \ln(1+2x)+\mathrm{Li}_2\left(-2x\right)\right]-\frac{x -1}{x }\,\mathrm{Li}_2\left(1-2x\right)\right\}-\frac{i M x}{8\pi} \,\Delta\sigma^{(2)}(x).\nn
\eea 
\end{subequations}
The fact that $\im f^{(2)} (\nu)=\nu\,\sigma^{(2)}(\nu)/4\pi $ and $\im g^{(2)} (\nu)=-\nu\,\Delta\sigma^{(2)}(\nu)/8\pi $ is in accordance with the optical theorem, cf.\ Eq.~(\ref{OptT}). 
Also, we have checked that the one-loop amplitudes indeed
satisfy the dispersion relations:
\begin{subequations}
\eqlab{DR}
\bea
\eqlab{fDR}
f^{(2)}(\nu)&=&\frac{2\nu^2}{\pi}\int_0^\infty \! \dd\nu' \,\frac{\im f^{(2)}(\nu')}{\nu'\left(\nu^{\prime \, 2}-\nu^2 - i0^+\right)},\\
g^{(2)}(\nu)&=&\frac{2\nu}{\pi}\int_0^\infty \! \dd\nu' \,\frac{\im g^{(2)}(\nu')}{\nu^{\prime \, 2}-\nu^2 - i0^+},
\eea
\end{subequations}
where the subtraction in \Eqref{fDR} corresponds to $f^{(2)}(0)=0$.

Now, we want to understand the low-energy expansion, and
thus the polarizability sum rules in spinor QED. In Ref.~\cite{Gryniuk:2015eza}, we already performed this exercise in scalar QED.
Expanding the real part of \Eqref{DR} around $\nu=0$,  we find
\begin{align}
 \frac{\alpha^2}{\pi M} \bigg(\frac{11+48 \ln \frac{2\nu}{M}}{18M^2 } \nu^2+\frac{7 (257+1140\ln \frac{2\nu}{M})}{450M^4}\nu ^4+\frac{68(107+672 \ln \frac{2\nu}{M})}{441M^6 }\nu ^6  + \ldots \bigg)
= \frac{1}{2\pi^2} \sum_{n=1}^\infty \nu^{2n} \int_0^\infty \!\dd\nu' \,\frac{\sigma^{(2)}(\nu')}{\nu^{\prime\, 2n}},\qquad\nn\\
\frac{\alpha^2}{\pi M} \bigg(\frac{ 37+60  \ln \frac{2\nu}{M}}{18 M^3}\nu ^3+\frac{64 (29+105  \ln \frac{2\nu}{M})}{225 M^5} \nu ^5
+\frac{18 (89+504 \ln \frac{2\nu}{M})}{49 M^7 } \nu ^7 +\ldots \bigg)= \frac{1}{4\pi^2} \sum_{n=1}^\infty \nu^{2n-1} \int_0^\infty \!\dd\nu' \,\frac{\Delta \sigma^{(2)}(\nu')}{\nu^{\prime\, 2n-1}}.\qquad\nn
\end{align}
Hence, the coefficients diverge in the infrared. However, there is an apparent mismatch:  
they are logarithmically divergent on one side
 and power divergent on the other. To match the sides exactly at
 each order of $\nu$, thus defining the sum rules for ``quasistatic'' 
 polarizabilities, we subtract all the power divergences on the
right-hand-side (rhs) and 
regularize both sides with the same infrared cutoff (equal to $\nu$):
\bea 
\eqlab{goodscalarLEX}
 \frac{\alpha^2}{\pi M} \bigg(\frac{11+48 \ln \frac{2\nu}{M}}{18M^2 } \nu^2+\frac{7 (257+1140\ln \frac{2\nu}{M})}{450M^4}\nu ^4 + \ldots \bigg)
&=& \frac{1}{2\pi^2} \sum_{n=1}^\infty \nu^{2n} 
\int_\nu^\infty \!\dd\nu' \,\frac{\sigma^{(2)}(\nu')-\sum\limits_{k=0}^{2(n-1)}\frac{1}{k!}\frac{\dd^k\sigma^{(2)}(\nu)}{\dd\nu^k}\Big|_{\nu=0}\,\nu^{\prime\, k}}{\nu^{\prime\, 2n}},\nn\\
\frac{\alpha^2}{\pi M} \bigg(\frac{ 37+60  \ln \frac{2\nu}{M}}{18 M^3}\nu ^3+\frac{64 (29+105  \ln \frac{2\nu}{M})}{225 M^5} \nu ^5 +\ldots \bigg)&=& \frac{1}{4\pi^2} \sum_{n=2}^\infty \nu^{2n-1} 
\int_\nu^\infty \!\dd\nu' \,\frac{\Delta\sigma^{(2)}(\nu')-\sum\limits_{k=0}^{2n-3}\frac{1}{k!}\frac{\dd^k\Delta\sigma^{(2)}(\nu)}{\dd\nu^k}\Big|_{\nu=0}\,\nu^{\prime\, k}}{\nu^{\prime\, 2n-1}}.\nn
\eea  
Both sides are  now identical at each order of $\nu$. This is nontrivial, at least for the analytic terms; the logs are fairly easily obtained from the nonregularized rhs of the low-energy expanded dispersion relation, cf.\ Ref.~\cite{Holstein:2005db}. Since the GDH sum rule only differs from zero starting from $O(\al^3)$, we omitted the $O(\nu)$ term in the last equation.

Extending these arguments to all orders in $\al$, we find that the proper
low-energy expansion for the ``elastic'' part of the amplitudes
reads as
\begin{subequations}
\bea
\eqlab{elLEX}
 f_{\mathrm{el}}(\nu)&=& -\,\frac{\al}{M} + \frac{1}{2\pi^2}\sum_{n=1}^\infty \nu^{2n} 
\!\int_\nu^\infty\! \dd\nu' \,\frac{\sigma(\nu')-\ol\sigma_n(\nu')}{\nu^{\prime\, 2n}}\,,\\
 g_{\mathrm{el}}(\nu)&=&  \frac{1}{4\pi^2}\sum_{n=1}^\infty \nu^{2n-1} 
\!\int_\nu^\infty\! \dd\nu' \,\frac{\Delta\sigma(\nu')-\ol{\Delta\sigma}_n(\nu')}{\nu^{\prime\, 2n-1}}\,,
\eea
\end{subequations}
\end{widetext}
where the bar denotes the infrared subtractions:
\begin{subequations}
\bea
\ol\sigma_n(\nu') &\equiv& \sum\limits_{k=0}^{2(n-1)}\frac{1}{k!}\frac{\dd^k\sigma(\nu)}{\dd\nu^k}\Big|_{\nu=0}\,\nu^{\prime\, k},\\
\ol{\Delta\sigma}_n(\nu')&\equiv&\begin{cases}
0&n=1\\
 \sum\limits_{k=0}^{2n-3}\frac{1}{k!}\frac{\dd^k\Delta\sigma(\nu)}{\dd\nu^k}\Big|_{\nu=0}\,\nu^{\prime\, k}&n>1.\qquad
\end{cases}
\eea
\end{subequations}
Accordingly, the elastic contributions to the polarizabilities are given by
\begin{subequations}
\bea
(\alpha_{E1}+\beta_{M1})_{\mathrm{el}} &=&\frac{1}{2\pi^2}\int_\nu^\infty \dd\nu' \,\frac{\sigma(\nu')-\sigma(0)}{\nu^{\prime\,2}},\\
(\gamma_0)_{\mathrm{el}} &=&-\frac{1}{4\pi^2} \int_\nu^\infty \dd\nu' \,\frac{\Delta\sigma(\nu')-\ol{\Delta\sigma}_2(0)}{\nu^{\prime\,3}},\\
(\bar{\gamma_0})_{\mathrm{el}} &=&-\frac{1}{4\pi^2} \int_\nu^\infty \dd\nu' \,\frac{\Delta\sigma(\nu')-\ol{\Delta\sigma}_3(0)}{\nu^{\prime\,5}}.\qquad\quad
\eea
\end{subequations}
In our one-loop spinor QED example, plugging in the tree-level cross sections from \Eqref{crossSectionTree}, we obtain
\begin{subequations}
\bea
(\alpha_{E1}+\beta_{M1})_{\mathrm{el}} &=& \frac{\alpha^2}{18\pi M^3}\left(11+48 \ln \frac{2\nu}{M}\right),\\
(\gamma_0)_{\mathrm{el}} &=&-\frac{\alpha^2}{18\pi M^4}\left(37+60  \ln \frac{2\nu}{M}\right),\\
(\bar{\gamma_0})_{\mathrm{el}} &=&-\frac{64\,\alpha^2}{225\pi M^6}\left(29+105  \ln \frac{2\nu}{M}\right),\qquad
\eea
\end{subequations}
which obviously matches the corresponding terms in the low-energy expansion of the tree-level amplitudes.

\newpage
  
\bibliographystyle{unsrtnat}

\end{document}